\documentclass[12pt,letter]{article}
\usepackage{fullpage, amsthm}
\newtheorem{proposition}{Proposition}
\newtheorem{problem}[proposition]{Problem}

\newtheorem{lemma}[proposition]{Lemma}

\title{Computational Complexity of the Monoid Frobenius Problem with Compressed Input}

\author{Jeffrey~Shallit, Zhi~Xu}
\date{}
\begin{document}

\maketitle

\begin{abstract}
The following problem is NP-hard: given a regular expression $E$,
decide if $E^*$ is not co-finite. This problem is also in PSPACE.
\end{abstract}

\section{Introduction}

Given $k$ positive integers $x_1,x_2,\cdots,x_k$ with
$\gcd(x_1,x_2,\cdots,x_k)=1$, the \emph{Frobenius Problem} is to
find the largest integer that cannot be represented as a
non-negative integer linear combination of the given integers. This
largest integer is called the \emph{Frobenius number} of the given
integers, and is denoted as $g(x_1,x_2,\cdots,x_k)$. One can refer
to \cite{Alfonsin2005} for a good survey on the Frobenius Problem.

There have been different generalizations of the Frobenius Problem
in the literature. One generalization of this problem to a free
monoid as follows: \cite{Kao2003} given a finite set $S$ of words on
a given alphabet, find the (length of) the longest word(s) not in
$S^*$ if $S^*$ is co-finite. One natural variation of this Monoid
Frobenius Problem is to give the input set of words $S$ in a
compressed way.

\begin{problem}\label{1}Given a regular expression $E$, find the length of
the longest word(s) not in $E^*$ if $E^*$ is co-finite.
\end{problem}

\section{NP-hardness Proof}
The NP-hardness of the following decision problem will be proved by
giving a polynomial reduction from 3SAT, using similar techniques in
the NP-completeness proof of star-free regular expression
inequivalence.

\begin{problem}\label{2}Given a star-free regular expression $E$,
decide if $E^*$ is not co-finite.
\end{problem}

Let $U=\{u_1,u_2,\cdots,u_n\}$ be a set of variables and
$C=\{c_1,c_2,\cdots,c_m\}$ be a set of clauses making up an
arbitrary instance of 3SAT. Without loss of generality, suppose each
variable appears in at least one clause. Then $n\leq 3m$. For each
clause $c_i$, a star-free regular expression $e_i=u_1^iu_2^i\cdots
u_n^i$ can be constructed, where $u_j^i=F$ if $u_j$ appears in
$c_i$, and $u_j^i=T$ if $\overline u_j$ appears in $c_i$, and
$u_j^i=(T+F)$ otherwise. Let $E'=e_1+e_2+\cdots+e_m,
E''=(T+F)(T+F)\cdots(T+F)=(T+F)^n, E=E'+E''(T+F)$. It is easy to
check this construction can be performed in polynomial time. It
remains to see that, in the above construction, the clauses $C$ are
satisfiable if and only if $E^*$ is not co-finite over the alphabet
$\Sigma=\{T,F\}$. The following lemma is required.

\begin{lemma}\cite{Kao2003}
     Suppose $S \subseteq \Sigma^m \cup \Sigma^n$, $0 < m < n$,
and $S^*$ is co-finite.  Then $\Sigma^m \subseteq S$.
\end{lemma}

Let $S$ be the set of words in $E$. Then $\Sigma^{n+1}\subseteq
S\subseteq\Sigma^n\cup\Sigma^{n+1}$, which follows by the
construction of $E$. If $C$ is satisfiable, then one can check that
$E'\not=E''$, so $\Sigma^n\not\subseteq S$. Therefore, $S^*$ cannot
be co-finite. If $S^*$ is not co-finite, since $\gcd(n,n+1)=1$, then
$\Sigma^n\not\subseteq S$, which leads to $E'\not=E''$. So, $C$ is
satisfiable. This finishes the NP-hardness proof of Problem \ref{2}.

As seen in the given polynomial reduction, restricting Problem
\ref{2} to the binary alphabet and/or demanding that the language of
$E$ consist of words of only two different lengths results in a
problem that is also NP-hard. The following problems can also easily
be seen to be NP-hard, as consequences of the NP-hardness of Problem
\ref{2}.

\begin{problem}\label{3} Given a regular expression $E$,
decide if $E^*$ is not co-finite.
\end{problem}

\begin{problem}\label{4} Given a NFA $M$,
decide if $S^*$ is not co-finite, where $S=L(M)$.
\end{problem}

\section{PSPACE Proof}
Both Problem \ref{3} and Problem \ref{4} are in PSPACE. A NPSPACE
algorithm for Problem \ref{3} will be presented, and the NPSPACE
proof for Problem \ref{4} follows in a straightforward manner.
Since NPSPACE = PSAPCE by Savitch's theorem, the result will follow.

Let $E$ be an arbitrary regular expression, and $t$ be the length of
$E$. An NFA $M$ can be constructed to accept $E^*$ with at most
$t+1$ states\cite{Holzer2002}. Let $n=2^{t+1}$. Then,
nondeterministically guess a word $w$ of length $i$ for $n\leq
i<2n$, and verify that it is rejected by $M$. $E^*$ is not co-finite
if and only if there exists a $w$ of length $i$ for $n\leq i<2n$
such that $M$ rejects $w$. To verify the word $w$ is rejected by
$M$, a boolean matrix of dimension  at most $t+1$ is stored to keep
track of reachability. The entry $(p,q)$ is true if and only if $q$
is reachable from $p$ on a given word. At the beginning, this matrix
is initialized as the identity matrix, for the word $\epsilon$. In each
step, the matrix is updated to process one guessed letter. This
requires only polynomial space.

To prove the correctness of this algorithm,  notice that an NFA can be
constructed to accept the language $E^*$ with at most $t+1$ states.
Hence a DFA $M'$ accepting $E^*$ has at most
$n=2^{t+1}$ states\cite{Yu1994}.  Interchanging final and nonfinal states,
the DFA $M''$
for $\overline{E^*}$ has at most $n$ states.  By a classical result
\cite{Hopcroft&Ullman}, $M''$ accepts an infinite language if and only
if it accepts a word of length $i$, $n \leq i < 2n$.  Hence $M'$'s
language is co-infinite if and only if it rejects some word of length
$i$, $n \leq i < 2n$.

\textbf{appendix}
\appendix This report is to be a part of a future paper.

\end{document}